Assessing randomness in case assignment: the case study of the Brazilian Supreme Court

Diego Marcondes[1], Cláudia Peixoto[2] and Julio Michael Stern[3]

April 19, 2018

[1] PhD Student, Department of Applied Mathematics, Institute of Mathematics and Statistics, University of São Paulo, Brazil. E-mail: dmarcondes@ime.usp.br.

[2] Professor, Department of Statistics, Institute of Mathematics and Statistics, University of São Paulo, Brazil. E-mail: claudiap@ime.usp.br.

[3] Professor, Department of Applied Mathematics, Institute of Mathematics and Statistics, University of São Paulo, Brazil. E-mail: jstern@ime.usp.br. I am grateful for the support received from FAPESP - the State of São Paulo Research Foundation (grants CEPID-CeMEAI 2013/07375-0 and CEPID-Shell-RCGI 2014/50279-4); and CNPq - the Brazilian National Counsel of Technological and Scientific Development (grant PQ 301206/2011-2).

We are grateful for the support received from IME-USP, the Institute of Mathematics and Statistics of the University of São Paulo; and ABJ, the Brazilian Jurimetrics Association. We are also grateful for advice and comments received from Adilson Simonis, Rafael Bassi Stern and Marcelo Guedes Nunes. We thank the *Supremo Tribunal Federal* (Brazilian Supreme Court) which made available the data analyzed in this article.

*κληρω νυν πεπαλασθε διαμπερες ος κε λαχησιν:*

*Let the lot be shaken for all of you, and see who is chosen*

Iliad, VII, 171.

מִדְיָנִים יַשְׁבִּית הַגּוֹרָל וּבֵין עֲצוּמִים יַפְרִיד

*Casting the dice puts judgement quarrels to rest and keeps powerful parts separated*

Proverbs 18:18.

**Abstract**

Sortition, i.e., random appointment for public duty, has been employed by societies throughout the years, especially for duties related to the judicial system, as a firewall designated to prevent illegitimate interference between parties in a legal case and agents of the legal system. In judicial systems of modern western countries, random procedures are mainly employed to select the jury, the court and/or the judge in charge of judging a legal case, so that they have a significant role in the course of a case. Therefore, these random procedures must comply with some principles, as statistical soundness; complete auditability; open-source programming; and procedural, cryptographical and computational security. Nevertheless, some of these principles are neglected by some random procedures in judicial systems, that are, in some cases, performed in secrecy and are not auditable by the involved parts. The assignment of cases in the Brazilian Supreme Court (*Supremo Tribunal Federal*) is an example of such procedures, for it is performed by a closed-source algorithm, unknown to the public and to the parts involved in the judicial cases, that allegedly assign the cases randomly to the justice chairs based on their caseload.

In this context, this article presents a review of how sortition has been employed historically by societies, and discusses how Mathematical Statistics may be applied to random procedures of the judicial system, as it has been applied for almost a century on clinical trials, for example. Based on this discussion, a statistical model for assessing randomness in case assignment is proposed and applied to the Brazilian Supreme Court in order to shed light on how this assignment process is performed by the closed-source algorithm. Guidelines for random procedures are outlined and topics for further researches presented.

## Introduction

The judicial systems of modern western states make use of randomization procedures, that is, important decisions in the course of a judicial case are made by rolling the dice. This may be perceived as an unnatural or counter-intuitive expedient. Why should the flip of a coin take part in casting the fate of a legal case? Why should we abdicate of making (best-informed) deterministic decisions? Next section addresses these questions, after a brief historical review. The following two sections make a statistical analysis designed to test the effective randomness of case assignment in the Brazilian Supreme Court. Last section gives our final remarks.

## Running Democracy and Rolling the Dice

The city-state of Athens was the undisputed champion in the use of sortition (random appointment) for managing its public affairs, especially in the period between the democratic reforms of Kleisthenes, in 507 BCE, and the fall of Athenian democracy, in 322 BCE; see Dowlen (2009), Hansen (1999), Headlam (1933) and Staveley (1972). While some specialized officers were elected based on specific expertise and experience, like military chiefs, water supply and treasury managers, religious priests etc., most of the public duties in Athens were carried out by citizens of the polis, that held short term positions appointed by a system based on district representation, sortition and frequent rotation.

Sortition was also used to constitute *dikasteria*, citizen's courts where *dike* (fair justice) was served. From a total of approximately 30,000 man recognized as qualified citizens (excluding women, slaves, foreigners and their descendants and other non-represented categories), 6,000 district representatives would be selected to the general assembly and, out of those, 501 individuals would be randomly chosen to constitute a *dikasterion*. Some specially important cases would require a larger court of justice, with up to 2,501 members! Moreover, in contrast to contemporary jurors, *dikastai* swore by the Heliastic Oath *not* to discuss the case in judgement among each other. In contrast, each *dikastes* should vote, by secret ballot, according to his individual and independent opinion, expressing his own best knowledge of the law and concern for the public good; see Hansen (1999, p.182).

The Hebrew bible uses randomization devices for several purposes, including ritualistic divination and judgement, and also as a practical tool for appeasement or fair distribution of goods and services; see for example Leviticus 16:8-10, Numbers 26:56, 1Chronicles 26:13, Proverbs 16:33 and Jonah 1:7. Biblical use of randomization was a strong argument for its practical adoption for political and legal procedures in Christian Europe, notwithstanding moral objections to gambling and philosophical concerns about voluntarily relaxing the grip of reason and relinquishing important decisions to fate.

Late Medieval and Renaissance Italy (12th to 17th century) saw many political experiments in the use of sortition to randomly appoint citizens for legislative and executive positions or legal functions, the two most important prototypes being the Florentine *scrutiny* and the Venetian *brevia*; see Najemy (1982) and Wolfson (1899). In England and the United States sortition was (and still is) mostly used for jury duty appointment, with mandatory random selection officially documented since the 17th century; see Dowlen (2009). Today, judicial systems of many western countries use randomization procedures to select the jury and/or the court in charge of judging a legal case.

All of the aforementioned historical examples include the appointment of jurors and judges by complex procedures involving a mixture of the following aspects: (1) Pre-selection of a pool of able or qualified candidates; (2) Voting or election by secret ballot; (3) Proportional representation constraints (by district, tribe or family, corporate guild, professional specialty, etc.); and, finally, at some point(s) in the appointment process, (4) Sortition, i.e., random selection among possible candidates.

Based on surviving original manuscripts and secondary literature, the references in this section give detailed accounts of *how* sortition was used in several historical examples. However, the original sources give us very little explanation of *why* sortition was used. Archaic religious texts see the ritual use of randomization devices as a doorway for the manifestation of divine will, but the aforementioned civilizations quickly realized that sortition also brought important practical advantages. Nevertheless, the articulation of good rational arguments for the use of sortition and randomization had to wait the development of Mathematical Statistics in the 20th century. This is the topic of our next section.

**Decoupling, Separation and Haphazardness**

Why do we randomize? Legal systems of modern societies use sortition as a firewall, a technological barrier designed to prevent spurious communication of vested interests or illegitimate interference between parties in a legal case and agents of the legal system. Its purpose is to warrant im-partial or non-partizan justice, literally, assure that no party (person, group of persons or social organizations), directly or indirectly involved or interested in a legal case, could illegally intervene or manipulate the due legal process.

Judges and jurors do not come to court as blank slates. They had a life history full of experiences that formed individual opinions, prejudices and idiosyncrasies. Hence, the opportunity given to a party to divert the case to a sympathetic judge or to select favorable jurors would constitute a virulent manipulation mechanism.

A similar situation is faced by statisticians and physicians when planning and conducting clinical trials, i.e., studies designed to test the performance of a newly proposed or alternative treatment for a given disease. For the sake of treatment comparison, patients taking part in the clinical trial have to be divided into a *control-group*, receiving an old treatment, a placebo, or no treatment whatsoever; and a *treatment-group*, receiving the newly proposed treatment. Giving the opportunity, participating patients would manifest their preferences, and try by all means possible to get their preferred choice of treatment. Wealthy and well-educated patients would be better informed, would make better choices, would be in better position to get their preferred treatment, but would also probably have better chances to overcome the disease and get well anyway. Hence, such *confounding effects* would ruin the very investigative purpose of the clinical trial; see Pearl (2004) and Stern (2008).

As well known by contemporary scientists, randomization is the key instrument used in statistics to overcome this difficult conundrum. In this context, *decoupling* refers to randomization techniques aiming to eliminate systematic vulnerabilities to uncontrolled influences received from or exerted by participating agents; see Lauretto et al. (2012, p.195). Charles Sanders Peirce (1839-1914), Joseph Jastrow (1863-1944) and Ronald Aylmer Fisher (1890-1962) introduced the key concept of randomization in Mathematical Statistics, as succinctly expressed by Judea Pearl in the following quotation; see also Peirce and Jastrow (1884), Fisher (1926), Fisher (1935), and Stern (2008).

> *Fisher's great insight was that connecting the new link to a random coin flip 'guarantees' that the link we wish to break is actually broken. The reason is that a random coin is assumed to be unaffected by anything we can measure on macroscopic level - including, of course, a patient's socioeconomic background.* (Pearl, 2000, p.348) also quoted in Stern (2008, p.59).

Pearl (2000) further develops the axiomatic study of related issues, and encodes the concepts of decoupling by randomization in the language of abstract inference diagrams, giving the all-important formal definition of *d-separation*. Furthermore, Dennis Lindley (1923-2013) uses the word *haphazardness* to clearly distinguish between the desired effect of *decoupling*, from randomness, the tool used to achieve it; see next quotation.

> *The Role of Randomization in Inference: We describe a possible allocation that the experimenter judges to be free of covariate interference as haphazard. Randomization may be a convenient way of producing a haphazard design. We argue that it is the haphazard nature, and not the randomization, that is important.* (Lindley, 1982, p.438-439), also quoted in Fossaluza et al. (2015, p.173).

Notwithstanding the important conceptual distinction highlighted by Lindley, randomization remains the most basic and fundamental tool used to achieve *decoupling* in modern statistics. Nevertheless, this conceptual distinction opens the possibility for improvement, by development of more efficient techniques used to render well-decoupled designs; see for example Stern (2008), Lauretto et al. (2012), Fossaluza et al. (2015), and Lauretto et al. (2017). Finally, the consistent use of randomization techniques as well as the compliance with the implied protocols can be investigated and audited with tools provided by Mathematical Statistics and computational data analysis; this is the goal of the next sections.

## Statistical Modelling of Case Assignments

A model to assess randomness in case assignment may be applied to the following abstract scenario. Suppose there is a courthouse constituted of *n* judges and that new cases arrive daily and must be assigned to one of them. Also, assume that it is of interest to study the mechanism of case assignment of the courthouse, in order to outline the existence or not of a bias towards assigning cases to a specific judge, under particular circumstances. To meet this goal, the probability of a new case being assigned to each judge, under given circumstances, may be modelled.

In order to fit such a model, a sample of case assignments is needed. The sample must contain the number of cases that were assigned to each judge daily for a fixed period of time. Furthermore, it must also present the circumstances under which such case assignments occurred. These circumstances are represented by variables that summarize the daily conditions of the courthouse regarding possible sources of bias on the assignment process. A reliable characterization of the daily conditions of the courthouse is essential, for it is desirable to model any source of bias on the case assignment process, so that any bias still present on the fitted model is due to randomness or an unknown source. As an example of known bias on case assignment, consider the scenario in which a case fits the related-case rule, so that it is assigned to a specific judge without partaking any random process. It is important that the circumstances under which this case was assigned be presented in the sample so that it may be incorporated to the model, as this is a known source of bias that alters the assignment probabilities of this specific case.

The model to be fitted is a Multinomial Logistic Regression, which is presented in details in the appendix. This regression models the probability of a case being assigned to a judge as a non-linear function of numerical variables that represent the conditions under which the case is assigned. Other kinds of known sources of bias, which may not be represented by numerical variables, may also be incorporated to the model.

## Case study of the Brazilian Supreme Court

The abstract model outlined on the preceding section is now fitted to a sample of case assignments of the Brazilian Supreme Court (*Supremo Tribunal Federal*). The sample contains the number of cases of each class assigned to each justice chair[4] of the court between February 28 2008 and July 10 2017. At the considered period of time, cases of 35 different classes were assigned to the 11 chairs of the court, although only 14 classes of cases are considered, as there were less than a thousand cases of the other classes assigned over the 10 year period, as they are somewhat rare when comparing to the top 14 classes. Considering only the assignments of the top 14 classes, the sample amounts to 22,720 assignments.[5]

Table 1 presents the number of cases of each class that were assigned during the considered period of time for each justice chair, and Figure 1 presents the proportion of the cases of the class Special Appeal with Aggravation that were assigned daily for each chair. It can be seen that there are some justice chairs that systematically receive a greater proportion of cases of this class for a period of time, although this may be explained by some kind of known source of bias and may not be evidence of lack of randomness in the case assignment process.

---

[4] The Brazilian Supreme Court is formed by justice chairs, each one having a minister. When a minister retires, its caseload is transferred to the newly appointed minister of his chair. Therefore, in order to follow the caseload of each minister over time, it is convenient to consider that the cases are assigned to the chair, not to the minister.

[5] One for each day and class. If in a day there were assigned cases of *m* different classes, then there were *m* random samples in this day.

Table 1. Number of cases assigned by class to each justice chair during the considered period of time[6].

| Class | Justice Chair | | | | | | | | | | | Total |
|---|---|---|---|---|---|---|---|---|---|---|---|---|
| | 1 | 2 | 3 | 4 | 5 | 6 | 7 | 8 | 9 | 10 | 11 | |
| Precautionary Action (AC) | 271 | 268 | 234 | 255 | 198 | 231 | 231 | 248 | 220 | 200 | 213 | 2569 |
| Originating Civil Action (ACO) | 214 | 172 | 204 | 213 | 143 | 180 | 156 | 189 | 146 | 190 | 196 | 2003 |
| Direct Unconstitutionality Action (ADI) | 212 | 159 | 147 | 230 | 157 | 160 | 134 | 197 | 171 | 156 | 177 | 1900 |
| Bill of Review (AI) | 16229 | 18353 | 14172 | 14644 | 9452 | 16226 | 17664 | 14605 | 14941 | 11445 | 10256 | 157987 |
| Special Appeal with Aggravation (ARE) | 37605 | 16848 | 20278 | 20184 | 19936 | 17789 | 9569 | 19988 | 13063 | 33778 | 19998 | 229036 |
| Habeas Corpus (HC) | 6535 | 3854 | 4529 | 4630 | 3436 | 4056 | 3349 | 4447 | 3708 | 5249 | 4249 | 48042 |
| Inquiry (Inq) | 216 | 166 | 183 | 200 | 172 | 194 | 153 | 198 | 216 | 189 | 106 | 1993 |
| Injunction Order (MI) | 672 | 601 | 592 | 607 | 398 | 485 | 553 | 573 | 451 | 449 | 560 | 5941 |
| Writ of Mandamus (MS) | 696 | 597 | 670 | 693 | 638 | 1347 | 426 | 661 | 671 | 379 | 642 | 7420 |
| Petition (Pet) | 198 | 169 | 187 | 181 | 367 | 127 | 137 | 214 | 276 | 331 | 126 | 2313 |
| Extraordinary Appeal (RE) | 17021 | 13352 | 11518 | 12240 | 8891 | 13730 | 11181 | 12031 | 10759 | 17891 | 7565 | 136179 |
| Habeas Corpus Appeal (RHC) | 599 | 398 | 472 | 451 | 393 | 403 | 326 | 459 | 276 | 543 | 439 | 4759 |
| Writ of Mandamus Appeal (RMS) | 169 | 145 | 168 | 153 | 117 | 125 | 137 | 161 | 96 | 101 | 147 | 1519 |
| Complaint (Rcl) | 3198 | 2007 | 2039 | 2537 | 1789 | 2018 | 1454 | 2190 | 1784 | 2702 | 2119 | 23837 |
| Total | 83835 | 57089 | 55393 | 57218 | 46087 | 57071 | 45470 | 56161 | 46778 | 73603 | 46793 | 625498 |

---

[6] The name of the classes were translated from Portuguese. Their abbreviations are kept as they are in Portuguese.

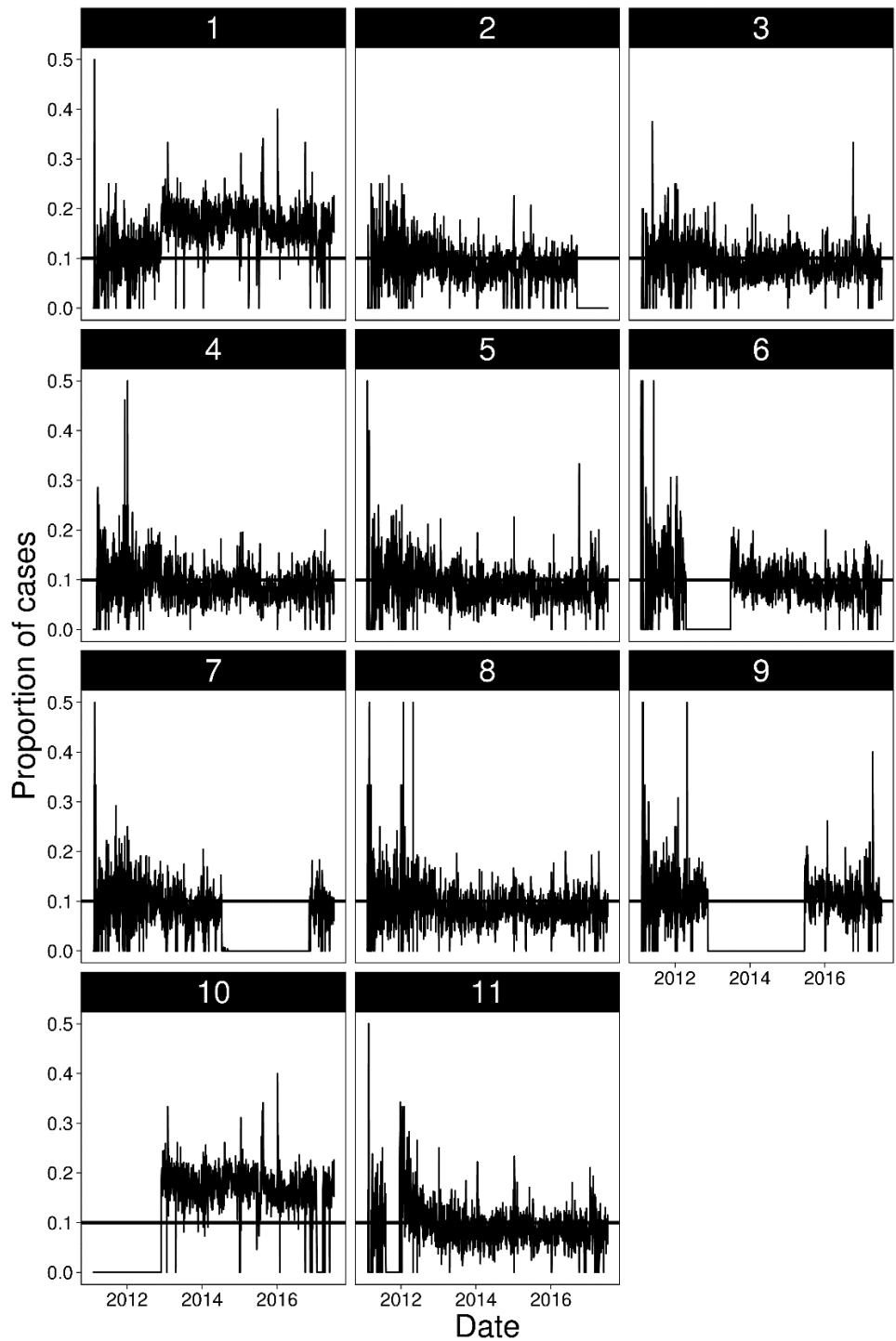

Figure 1: Proportion of cases of the class Special Appeal with Aggravation assigned daily to each chair. The dotted line represents the 0.10 proportion, which is the proportion of cases that is expected to be assigned to each chair if the assignment process is performed randomly. This class of case was created in 2011 and the period of time in which a chair systematically receives no case refers to when the chair was empty.

The case assignment process of the Brazilian Supreme Court has some known sources of bias that must be controlled. Indeed, although 11 justice chairs form the court, at least one of them is unavailable for receiving new cases, as one of the ministers is the court president and, therefore, occupies the presidency chair and does not receive new cases from the random assignment process[7]. Furthermore, there may be other empty chairs, for when a minister retires, his chair remains empty for a period of time, until a new minister is appointed by the President of the Republic and approved by the senate. Indeed, these are known sources of bias that are included in the model as a missing data mechanism and not as numerical variables that describe the circumstances of the case assignment. In Statistics, a sample with missing data is such that some values of its variables are missing, i.e., are not known; see Little and Rubin (2014). In the present case, it is not known how many cases would have been assigned to the chair of the president minister if it was available for receiving cases in a given day, what characterizes a missing data situation. The missing data mechanism in the case at hand is structured, i.e., the mechanism that causes the missing data is known, and, thus, can be modelled. Introducing the missing data factor into the model, it becomes a Multinomial Logistic Regression Model with Missing Data, for which more details are given in the appendix.

Another source of bias is the existence of the related-case rule, by which some cases are not randomly assigned, but are directly assigned to a specific chair that is handling related cases. However, the available sample does not distinguish the cases that were randomly assigned from the cases that were not and, therefore, this source of bias cannot be incorporated to the model.

---

[7] The court president receives cases that are exclusive to the presidency chair. Therefore, the president minister does not partake on the random assignment process.

The model for the Brazilian Supreme Court considers only one variable describing the circumstances under which each case assignment was performed. The variable considered in our analysis is the proportion of cases of each class that has been assigned to each chair since 2001 to the day of the assignment. In fact, at each day, this proportion is updated by the cases assigned at the previous day. It is important to consider this variable for the court claims that the assignment is performed based on this quantity, so that chairs with high proportion have lower probability of being assigned a new case of the class. However, the court does not disclose how the assignment algorithm takes into account this proportion when assigning cases. Therefore, the probability of a chair receiving a new case of a given class is modelled as a non-linear function of the proportion of cases of the class assigned to the chair since 2001 to the day of the assignment. This non-linear function differs from one chair to another and from one class to another.

Figure 2 presents the confidence intervals for the probabilities of a new case of a given class being assigned to each chair, under the circumstance that each justice chair has received 1/11 of the cases of the class since 2001 to the assignment date. These intervals were estimated by the Multinomial Logistic Regression detailed in the appendix. From the estimated intervals, we see that, in all classes, the probability of a chair receiving new cases is always less than 0.30, so that there are not evidences of great biases on the assignment process. Nevertheless, apart from class AC, in which all intervals contain[8] 1/11, for all other classes we reject that the probability is the same for all chairs, i.e., the difference between the probabilities of assignment are statistically significant. However, the difference is not *practically* significant on most of the classes, as the probabilities are homogeneous, though different (see classes AI, Inq and RHC for example).

On the other hand, there are classes in which some chairs have a higher probability of receiving a new case (see classes ADI, ARE, MS, Pet and RMS for example). In these classes, even though all chairs have the same proportion of cases, some of them have a higher probability of receiving a new case, what points to the existence of some bias on the assignment process. However, this bias may not be part of the assignment process itself, but be rather caused by some variable that is not being taken into account by our model (as the related-case rule, for example). Therefore, though we see that the probability of some chairs receiving a new case is greater, we cannot conclude that the process is biased (at least not without further studies taking into account other factors that may bias the process).

---

[8] So that we do not reject the hypothesis that the probability of a new case being assigned to a chair is equal for all chairs.

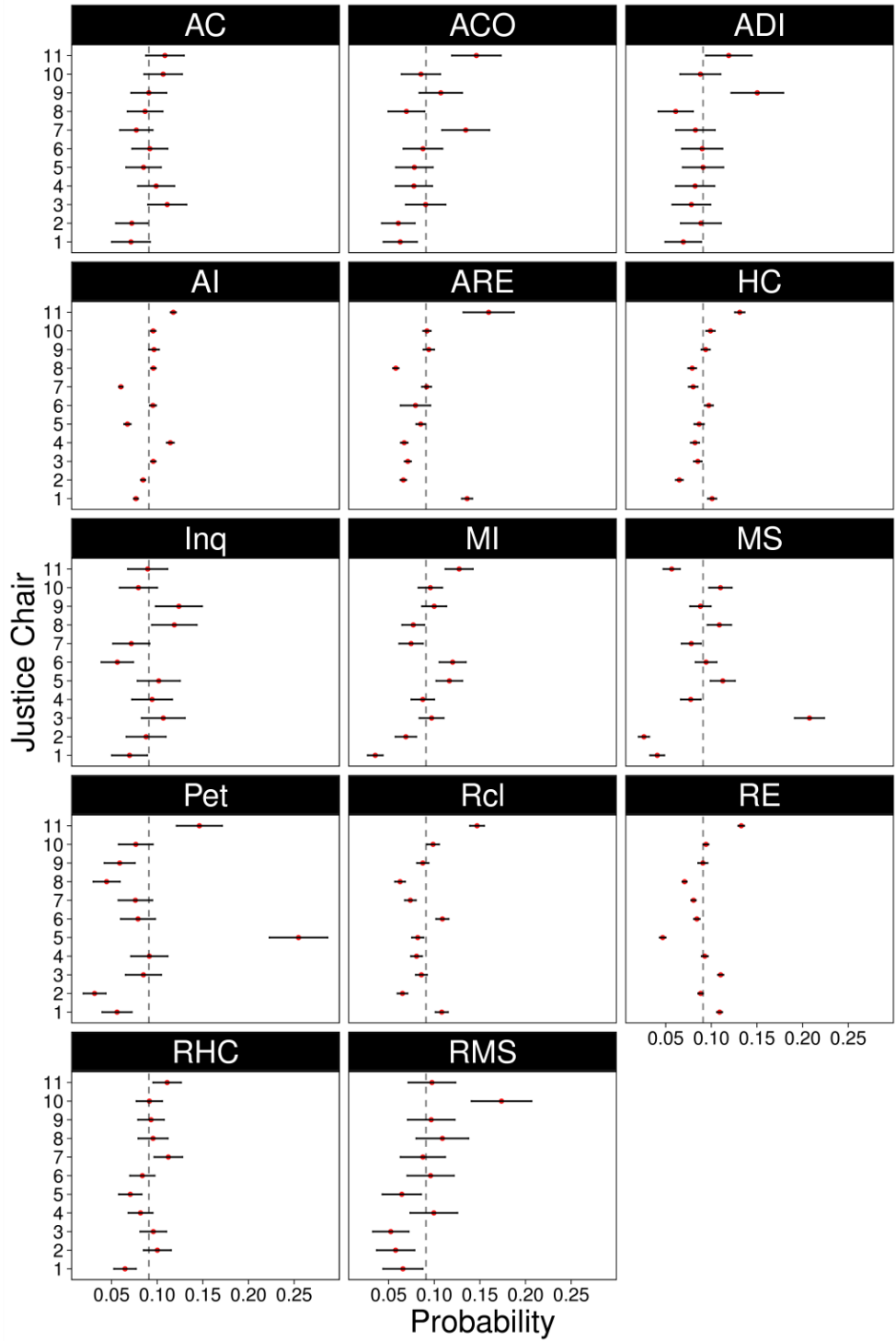

Figure 2: Confidence intervals for the probabilities of a new case of a given class being assigned to each justice chair, under the circumstance that each chair has received *1/11* of the cases of the class since 2001 to the assignment date, estimated by the Multinomial Logistic Regression. The confidence of the intervals is 99% and they are corrected by the Bonferroni method (Neter et al., 1990, p.164).

**Final Remarks and Future Research**

What we can conclude and what we cannot

The statistical analysis in the preceding sections give us a global assessment of the allocation process. Although we detect deviations from expected statistics, indicating the existence of systematic biases on the random allocation of judicial cases, we cannot make any conclusions about the fairness or appropriate randomization process of any individual judicial case. Such a conclusion cannot be reached by post-hoc statistical analysis of historical data. Furthermore, the systematic biases may be created by sources that, tough unknown on the dataset at hand, are known to the court, as the cases that fit the related-case rule, for example.

Nevertheless, given the juridical and social importance of this theme, we believe that it is important to develop software and procedures for randomization in compliance with the following desiderata: (a) Statistical soundness; (b) Procedural, cryptographical and computational security; (c) Complete auditability; (c) Open-source programming; (d) Multiple hardware platform and operating system implementation; (d) User friendliness and transparency; (e) Flexibility and adaptability for the needs and requirements of multiple application areas (like, for example, clinical trials, selection of jury or judges in legal proceedings, and draft lotteries). This is an important topic of further research.

Final remarks

Stern (2008), Lauretto et al. (2012), Fossaluza et al. (2015) and Lauretto et al. (2017) reconsider the notion of decoupling, as used in statistics, with the purpose of allowing multi-objective optimization of (approximate) separation criteria together with other desired goals, like (bounded) proportional representation criteria. Originally developed in the context of clinical trials, these methods could be easily adapted to applications in the legal system.

In contrast to other jurisdictions, Brazilian courts of appeal allow formation of review panels by spontaneous aggregation of judges. Indubitably, this factor has important impacts in the properties of the appeal process concerning statistical randomization. The issues addressed in the previous paragraphs, and several correlated matters constitute interesting topics for further research.

**Appendix: Technical details**

In this appendix, we present the details of the Multinomial Logistic Regression that is applied to the Brazilian Supreme Court. This regression model aims to model the logarithm of the odds in favor of a case being assigned to a justice chair over a so-called reference chair under given circumstances. The mathematical structure of the model imposes that a reference chair is fixed so that the probability of a case being assigned to a chair is compared with the probability of this same case being assigned to the reference chair through the respective odds. The logarithm of the odds is then modelled as a linear function of the numerical variables that represent the courthouse circumstances at the moment of the assignment. For more details about Logistic Regression models see Hilbe (2009).

Let $J$ be the variable that indicates to which justice chair a given case is to be assigned. This variable takes values in $\{1, \dots, n\}$, as each one of the $n$ chairs may be represented by a number. Also, let $(x_{1,j}, \dots, x_{k,j})$ be a set of numerical variables that represent the circumstances of the chair $j$ when the assignment occurred, and define $(v_1, \dots, v_n)$ as $v_j = 1$ if the chair $j \in \{1, \dots, n\}$ is available for receiving cases, and zero otherwise. Note that $v_j$ introduces a structured missing data mechanism into the model, i.e., the mechanism that causes a chair to be empty. Considering that the reference chair is the number 1 and that $v_1 = 1$, i.e., the reference chair is always available for receiving cases, the Multinomial Logistic Regression with Missing Data may be written as

$$\begin{cases} \log\left(\dfrac{P\left(J = j \mid (x_{j,1}, \dots, x_{j,k}), (1, v_2, \dots, v_n)\right)}{P\left(J = 1 \mid (x_{j,1}, \dots, x_{j,k}), (1, v_2, \dots, v_n)\right)}\right) = \beta_0^{(j)} + \beta_1^{(j)} x_{j,1} + \cdots + \beta_k^{(j)} x_{j,k} & if\ v_j = 1 \\ P\left(J = j \mid (x_{j,1}, \dots, x_{j,k}), (1, v_2, \dots, v_n)\right) = 0 & if\ v_j = 0 \end{cases} \quad (1)$$

for $j = 2, \ldots, n$, in which $P$ represents probability. Observe that in model (1) each chair has its own parameters, though we can restrict this model making some parameters equal for all chairs. In addition, for simplification, we consider that all chairs have the same number $k$ of parameters, what is not necessary.

In order to estimate the parameters $\beta_0^{(j)}, \ldots, \beta_k^{(j)}, j \in \{2, \ldots, n\}$, of the model by the Maximum Likelihood Method it is necessary to have a sample $(Y, V, X)$ of size $m$ given by

$$V = \left((v_{1,1}, \ldots, v_{1,n}), \cdots, (v_{m,1}, \ldots, v_{m,n})\right)$$

$$Y = \left((y_{1,1}, \ldots, y_{1,n}), \cdots, (y_{m,1}, \ldots, y_{m,n})\right)$$

$$X = \left((\mathbf{x}_{1,1}, \ldots, \mathbf{x}_{1,n}), \cdots, (\mathbf{x}_{m,1}, \ldots, \mathbf{x}_{m,n})\right)$$

$$\mathbf{x}_{s,j} = (x_{s,j,1}, \ldots, x_{s,j,k}); s = 1, \ldots, m; j = 1, \ldots, n$$

in which $y_{s,j}$ is the number of cases assigned to chair $j$ at assignment $s$; $v_{s,j}$ equals one if chair $j$ was available for receiving cases at assignment $s$ and zero otherwise; and $\mathbf{x}_{s,j}$ are the numerical variables that describe the circumstances of chair $j$ at assignment $s$. The likelihood of the model may be written as

$$L(\boldsymbol{\beta}) \propto \prod_{s=1}^{m} \prod_{j=1}^{n} \left( \frac{v_{s,j} \exp\left(\beta_0^{(j)} + \beta_1^{(j)} x_{s,j,1} + \cdots + \beta_k^{(j)} x_{s,j,k}\right)}{1 + \sum_{l=2}^{n} v_{s,l} \exp\left(\beta_0^{(l)} + \beta_1^{(l)} x_{s,l,1} + \cdots + \beta_k^{(l)} x_{s,l,k}\right)} \right)^{y_{s,j} \times v_{s,j}}$$

with the convention that $\beta_0^{(1)} = \beta_1^{(1)} = \cdots = \beta_k^{(1)} = 0$. The estimates $\hat{\beta}_0^{(j)}, \ldots, \hat{\beta}_k^{(j)}, j \in \{2, \ldots, n\}$, of the parameters are obtained by maximizing the likelihood presented above, and the probability of a new case being assigned to chair $j$, under new circumstances $\mathbf{X}_{m+1} = (\mathbf{x}_{m+1,1}, \ldots, \mathbf{x}_{m+1,n})$ and $\mathbf{V}_{m+1} = (v_{m+1,1}, \ldots, v_{m+1,n})$ is estimated by

$$P(J = j\,|\mathbf{X}_{m+1}, \mathbf{V}_{m+1}) = \frac{v_{m+1,j}\exp\left(\hat{\beta}_0^{(j)} + \hat{\beta}_1^{(j)} x_{m+1,j,1} + \cdots + \hat{\beta}_k^{(j)} x_{m+1,j,k}\right)}{v_{m+1,1} + \sum_{l=2}^n v_{m+1,l}\exp\left(\hat{\beta}_0^{(l)} + \hat{\beta}_1^{(l)} x_{m+1,l,1} + \cdots + \hat{\beta}_k^{(l)} x_{m+1,l,k}\right)}$$

again with the convention that $\hat{\beta}_0^{(1)} = \hat{\beta}_1^{(1)} = \cdots = \hat{\beta}_k^{(1)} = 0$. Confidence intervals for these probabilities may also be obtained applying the properties of Maximum Likelihood Estimators (Rao, 1973, Chapter 5).

We will consider six distinct models for the Brazilian Supreme Court that are particular forms of (1). In the first model, we suppose that the logit (1) depends on the class of the case and on the proportion of cases assigned to the chair since 2001, and that both the effect of class and proportion depend on the chair:

$$\begin{cases} \log\left(\dfrac{P\big(J=j\,\big|(x_{j,1},\ldots,x_{j,15}),(1,v_2,\ldots,v_{11})\big)}{P\big(J=1\,\big|(x_{j,1},\ldots,x_{j,15}),(1,v_2,\ldots,v_{11})\big)}\right) = \beta_1^{(j)} x_{j,1} + \cdots + \beta_{15}^{(j)} x_{j,15} & \text{if } v_j = 1 \\ P\big(J=j\,\big|(x_{j,1},\ldots,x_{j,15}),(1,v_2,\ldots,v_{11})\big) = 0 & \text{if } v_j = 0 \end{cases}$$

for $j = 2, \ldots, 11$, in which $x_{j,i} = 1, i \in \{1, \ldots, 14\}$,[9] if the class of the process is the $i-th$ and zero otherwise, and $x_{j,15}$ is the proportion of cases of the class assigned to chair $j$ since 2001. There are 150 parameters in this model, one for each class and chair (140) and one for each chair that refers to the proportion (10). In this model, not only the probability of a case being assigned to a given chair depends on the proportion of cases of this class assigned to it since 2001 and the class of the case, but also the dependence depends on the chair, i.e., the effect of the proportion and the class on the probability is different for each chair.

---

[9] We represent the classes by numbers as it makes the notation clearer. We may number the classes in an arbitrary order, without loss of generality.

In model 2, we suppose that the logit (1) depends on the proportion and on the class, but the effect of the class is the same for all chairs, while the effect of the proportion depends on the chair:

$$\begin{cases} \log\left(\frac{P(J=j|(x_{j,1},\dots,x_{j,15}),(1,v_2,\dots,v_{11}))}{P(J=1|(x_{j,1},\dots,x_{j,15}),(1,v_2,\dots,v_{11}))}\right) = \beta_1 x_{j,1} + \cdots + \beta_{14} x_{j,14} + \beta_{15}^{(j)} x_{j,15} & \text{if } v_j = 1 \\ P(J=j|(x_{j,1},\dots,x_{j,15}),(1,v_2,\dots,v_{11})) = 0 & \text{if } v_j = 0 \end{cases}$$

for $j = 2, \dots, 11$, in which $x_{j,i} = 1, i \in \{1, \dots, 14\}$, if the class of the process is the $i-th$ and zero otherwise, and $x_{j,15}$ is the proportion of cases of the class assigned to chair $j$ since 2001. Note that in this model we restrict the parameters of model 1 to $\beta_i = \beta_i^{(2)} = \cdots = \beta_i^{(11)}$ for all $i \in \{1, \dots, 14\}$. Model 2 has only 24 parameters, one for each class (14), and one for each chair (10), which refers to the proportion.

Model 3 is a reduction of model 2, which supposes that the probability depends on the class and the proportion, but not on the chair, i.e., the effect of class and proportion is the same for all chairs:

$$\begin{cases} \log\left(\frac{P(J=j|(x_{j,1},\dots,x_{j,15}),(1,v_2,\dots,v_{11}))}{P(J=1|(x_{j,1},\dots,x_{j,15}),(1,v_2,\dots,v_{11}))}\right) = \beta_1 x_{j,1} + \cdots + \beta_{15} x_{j,15} & \text{if } v_j = 1 \\ P(J=j|(x_{j,1},\dots,x_{j,15}),(1,v_2,\dots,v_{11})) = 0 & \text{if } v_j = 0 \end{cases}$$

for $j = 2, \dots, 11$, in which $x_{j,i} = 1, i \in \{1, \dots, 14\}$, if the class of the process is the $i-th$ and zero otherwise, and $x_{j,15}$ is the proportion of cases of the class assigned to chair $j$ since 2001.

This rather simple model has only 15 parameters, one for the proportion and one for each class.

In model 4, we suppose that the probability does not depend on the class, but only on the proportion and on the chair:

$$\begin{cases} \log\left(\frac{P(J=j|(x_{j,1}),(1,v_2,\ldots,v_{11}))}{P(J=1|(x_{j,1}),(1,v_2,\ldots,v_{11}))}\right) = \beta_0^{(j)} + \beta_1^{(j)} x_{j,1} & \text{if } v_j = 1 \\ P(J=j|(x_{j,1}),(1,v_2,\ldots,v_{11})) = 0 & \text{if } v_j = 0 \end{cases}$$

for $j = 2, \ldots, 11$, in which $x_{j,1}$ is the proportion of cases (of a given class) assigned to chair $j$ since 2001. This model has only 20 parameters, two for each chair, and supposes that the case class does not affect the probability of assignment.

Model 5 is analogous to model 4, as we suppose that the probability does not depend on the proportion, but only on the class and the chair:

$$\begin{cases} \log\left(\frac{P(J=j|(x_{j,1},\ldots,x_{j,14}),(1,v_2,\ldots,v_{11}))}{P(J=1|(x_{j,1},\ldots,x_{j,14}),(1,v_2,\ldots,v_{11}))}\right) = \beta_1^{(j)} x_{j,1} + \cdots + \beta_{14}^{(j)} x_{j,14} & \text{if } v_j = 1 \\ P(J=j|(x_{j,1},\ldots,x_{j,14}),(1,v_2,\ldots,v_{11})) = 0 & \text{if } v_j = 0 \end{cases}$$

for $j = 2, \ldots, 11$, in which $x_{j,i} = 1, i \in \{1,\ldots,14\}$, if the class of the process is the $i-th$ and zero otherwise. This model has 140 parameters, one for each class and chair, and supposes that the assignment is not affected by the proportion, but only by the chair and the class.

Finally, model 6 supposes that the probability depends only on the proportion:

$$\begin{cases} \log\left(\frac{P(J=j|(x_{j,1}),(1,v_2,\ldots,v_{11}))}{P(J=1|(x_{j,1}),(1,v_2,\ldots,v_{11}))}\right) = \beta_0 + \beta_1 x_{j,1} & \text{if } v_j = 1 \\ P(J=j|(x_{j,1}),(1,v_2,\ldots,v_{11})) = 0 & \text{if } v_j = 0 \end{cases}$$

for $j = 2, \ldots, 11$, in which $x_{j,1}$ is the proportion of cases (of a given class) assigned to chair $j$ since 2001. This model supposes that the probability does not depend on the chair nor the class, but only on the proportion of cases assigned since 2001, and has only 2 parameters.

In order to properly choose a model for the Brazilian Supreme Court, we performed Likelihood Ratio Tests to test hypothesis of interest. The models' log-likelihoods, the hypothesis being tested and the p-value of the tests are presented in Table 2. The hypothesis always refer to model 1, i.e., suppose that model 1 is the true model, but the assumptions of the hypothesis. As we reject all hypothesis in Table 2, we choose model 1 as the one that better represents the case assignment process of the Brazilian Supreme Court. This is the model used to estimate the probabilities and confidence intervals in Figure 2.

| Model | ll | df | Chi-squared | p-value | Hypothesis |
|---|---|---|---|---|---|
| 1 | -1431800.65 | - | - | - | - |
| 2 | -1447745.67 | 126.00 | 31890.05 | < 0.0001 | The effect of class is the same on all chambers |
| 3 | -1452492.20 | 135.00 | 41383.10 | < 0.0001 | The effect of class and proportion is the same on all chambers |
| 4 | -1436513.10 | 130.00 | 9424.91 | < 0.0001 | There is no effect of class |
| 5 | -1433758.38 | 10.00 | 3915.46 | < 0.0001 | There is no effect of proportion |
| 6 | -1454554.17 | 148.00 | 45507.05 | < 0.0001 | There is no effect of class and the effect of proportion is the same on all chambers |